\documentstyle[aps,prl,multicol,epsf,epsfig]{revtex}

\def\CC{{\rm\kern.24em \vrule width.04em height1.46ex depth-.07ex
\kern-.30em C}}
\def\P{{\rm I\kern-.25em P}}
\def\RR{{\rm
         \vrule width.04em height1.58ex depth-.0ex
         \kern-.04em R}}

\def\RR{{\rm\kern.24em \vrule width.04em height1.46ex depth-.07ex
\kern-.30em R}}
\def\P{{\rm I\kern-.25em P}}
\def\RR{{\rm
         \vrule width.04em height1.58ex depth-.0ex
         \kern-.04em R}}

\newcommand{\be}{\begin{equation}}
\newcommand{\ee}{\end{equation}}
\newcommand{\bq}{\begin{eqnarray}}
\newcommand{\eq}{\end{eqnarray}}

\newcommand{\no}{\nonumber\\}

\newcommand{\p}{\partial}
\newcommand{\la}{\lambda}

\newcommand{\al}{\alpha}

\newcommand{\ep}{\epsilon}

\newcommand{\th}{\theta}

\newcommand{\lan}{\langle}
\newcommand{\ran}{\rangle}

\begin{document}
%\baselinestretch{2}
%\draft
\title{Topological Features in Ion Trap Holonomic Computation}
\author{Jiannis Pachos}
\address{ 
Max-Planck-Institut f\"ur Quantenoptik, D-85748 Garching, Germany
}
\maketitle
\begin{abstract}
{
Topological features in quantum computing provide controllability and noise
error avoidance in the performance of logical gates. While such resilience
is favored in the manipulation of quantum systems, it is very hard to identify
topological features in nature. This paper proposes a scheme where
holonomic quantum gates have intrinsic topological features. An ion trap is
employed where the vibrational modes of the ions are coherently manipulated 
with lasers in an adiabatic cyclic way producing geometrical holonomic
gates. A crucial ingredient of the manipulation procedures is squeezing of the
vibrational modes, which effectively suppresses exponentially any undesired
fluctuations of the laser amplitudes, thus making the gates resilient to control
errors. 
}
\end{abstract}
\date{today}
\pacs{PACS numbers:}
\maketitle
%\begin{multicols}{2}
%\narrowtext
\widetext
\section{Introduction}

In order to perform quantum computation (QC) effectively, it is
necessary to protect our system from control and noise errors. While
such an issue is easily dealt with in classical computation, the
quantum regime involves intrinsic characteristics which make quantum
error avoidance or error correction an important issue for QC. The
possible errors, due to the smallness of the system under
consideration, include the questionable viability of the
approximations considered and the controllability of the experimental
setup used to reproduce the desired ideal regimes, as well as thermal
noise and spontaneous emission. Two main approaches have been
developed in order to overcome these problems. The use of
error-correcting algorithms \cite{Duan}, which theoretically enable QC
to be carried out even in the presence of quantum noise, and
error-avoiding schemes, which produce a computational
(decoherence-free \cite{Zanardi,Lidar,Beige}) subspace of the whole
Hilbert space of the system, that is subject to minimal error
fluctuations. Such schemes include topological quantum computation
\cite{KIT,FREE,LLO}, the use of cooling, adiabaticity or the Zeno
effect \cite{Beige}. Both strategies function with resources
additional to those used for ideal encoding and processing quantum
algorithms. This extension of resources, properly treated allows the
overall control or noise error to be decreased \cite{Preskill}. There
is an analogy between the algorithmic strategies and these engineered
setups. Loosely speaking, one may regard the error-correcting
algorithm as ``simulating'' a corresponding error-avoiding physical
process.

In the literature there has been much interest in topological quantum
computation. Ideally, it constitutes within a certain control
procedure error-free computation. Although the proposed schemes have
hitherto been hard to realize experimentally, the concept is worth
investigating. Our particular aim here is to identify some topological
features from the geometrical ones \cite{AK,hol1} appearing in the
ionic setup. Let us focus on the case of small errors, called here
$\varepsilon$. While topological QC has no dependence in
$\varepsilon$ in all orders, our aim with geometrical QC is to
``neutralize'' our gates only up to a finite order in $\varepsilon$.
As an example, it is easy to visualize in our setup cancellation of
the first order in statistical errors in the variables the
experimenter is controlling. The parameters $\Sigma$ of the gates
presented here are engineered functions of the experimental parameters
in contrast to dynamic QC, where the gate parameters are merely linear
functions. In particular, each $\Sigma$ can be interpreted as the area
of a contour lying on a specific surface. This area is resilient to
the first order in statistical fluctuations of its border which the
experimentalist is traversing \cite{AK,Ellinas}. The weight factor in
the surface integral $\Sigma$, will further govern the robustness or
weakness of the gates.

Let us initially give in an abstract way the main ideas of holonomic quantum
computation (HQC) \cite{hol1,hol2,JAMS,Fujii}. Holonomies are a
generalization of the geometrical Berry 
phases \cite{SHWI} to the case of a multiple, e.g. $2^n$-fold,
degenerate Hamiltonian $H_0$.
The quantum information is encoded in an $2^n$-dimensional, degenerate
eigenspace $\cal C$ of $H_0$, with eigenvalue $E_0$, which is usually taken
to be the lowest ground energy. The operator
$H_0$ is considered to belong to the family ${\cal F}=\{H_\sigma={\cal
U}(\sigma)\,H_0\, 
{\cal U}^\dagger(\sigma);\sigma\in{\cal M}\}$ of Hamiltonians unitarily
(${\cal U}^\dagger (\sigma) ={\cal U}^{-1}(\sigma)$) equivalent to and therefore
iso-spectral with $H_0$, where $H_0=H_{\sigma_0}$ for some $\sigma_0 \in {\cal M}$.
The $\sigma$'s represent the classical ``control'' parameters used
to manipulate the encoded states $|\psi\rangle\in{\cal C}$.
Let $C$ be a {\em loop} in the control manifold $\cal M$.
When $C$ is slowly traversed, then the evolution is adiabatic and no
population is transferred between different energy levels.
If $|\psi\rangle_{in}\in{\cal C}$ is an initial state in the degenerate 
eigenspace, at the end of the loop it becomes
$ |\psi\rangle _{out}=e^{-i\,E_0\,T}\, \Gamma_{A}(C) \,|\psi\rangle_{in}$,
which still belongs in the same degenerate subspace. 
The first factor is just an overall dynamical phase which is omitted in the following 
by redefining the energy levels, i.e. by taking $E_0=0$. 
The second contribution is the holonomy $\Gamma_{A}(C)\in U(2^n)$ 
and is a result of the non-trivial topology of the {\em bundle} of eigenspaces 
over $\cal M$. By introducing the Wilczek-Zee connection~\cite{WIZE}
\begin{equation}
A_{\sigma_i}^{\bar \nu \nu}:= \langle \bar \nu |{\cal U}^\dagger(\sigma)
\,{\partial \over \partial\sigma_i}\,{\cal U}(\sigma)|\nu \rangle \,\, ,
\label{conn}
\end{equation}
where $A_{\sigma_i}^{\bar \nu \nu}$ is the $(\bar \nu, \nu)$ matrix
element of the $\sigma_i$ component of the connection,
one finds $\Gamma_{A}(C) ={\bf{P}}\exp \int_C A$, \cite{SHWI}, where
${\bf{P}}$ denotes path ordering. 
The set $H(A):=\{\Gamma_{A}(C);\forall C\in{\cal M}\} \subset U(2^n)$
is known as the holonomy group 
\cite{NAK}. In the case where it coincides with the whole unitary
group $U(2^n)$ the connection $A$ is called {\em irreducible} \cite{hol1}.
The transformations $\Gamma_A(C)$ for suitable $C$'s can be used as logical gates for HQC.

Let us apply these ideas to the ion traps. First of all we need to identify
degeneracy between two states. This is provided by the lowest
eigenstates of a Jaynes-Cummings Hamiltonian describing two internal
ionic levels ``dressed'' with the vibrational modes of the ion. Manipulating
the vibrational modes in an adiabatic cyclic fashion generates a holonomic operator
which evolves the degenerate dressed states. These control
procedures, which include one and two-mode displacing and squeezing,
can be performed by employing lasers which make coherent manipulations
between the Fock states of the harmonic modes. Let us look more explicitly at
these manipulation and the conditions that need to be satisfied in order to obtain
degeneracy. 

\section{Squeezing and Displacing of Vibronic oscillating modes}

In the following, the aim is to produce squeezing and displacing
operations on the oscillating mode of a trapped ion. This is produced with lasers
which couple the vibrational mode with two internal energy levels of the atom. In
particular, two standing-wave lasers are to be used for squeezing.
With their appropriate frequencies they will produce coherences between the
vibrational states $|n-1\ran$ and $|n+1\ran$ which simulate a squeezing
operation. In addition, a traveling wave will be used to insert a
displacing term in the overall Hamiltonian. Let us describe briefly the setup
first introduced in \cite{Cirac}.
\begin{center}
\begin{figure}[ht]
\centerline{
\put(105,78){$-\nu$}
\put(-20,88){$\omega_2$}
\put(-20,0){$\omega_1$}
\put(105,93){$+\nu$}
\put(0,40){$\Omega_a$}
\put(80,40){$\Omega_b$}
\put(50,40){$\Omega_0$}
  \epsffile{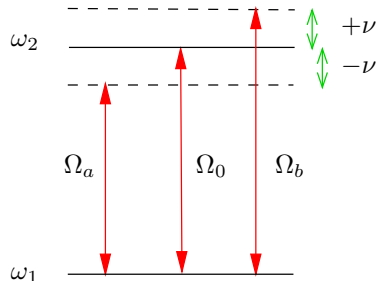}
}
  \caption[contour]{\label{ionn}
Atomic levels with vibronic modes and laser detunings.
           }
\end{figure}
\end{center}
A two-level trapped ion assumed to be located at a common node of
two standing-wave laser fields with the frequencies located symmetrically
about a carrier frequency $\omega_0$ at which we put an additional
traveling-wave field \cite{Cirac}. Using traveling waves for all lasers will
produce the same result \cite{Guridi}, with simplifications in the
experimental setup, even though it might make it more difficult to meet our
theoretical limits \cite{Jonathan}. The resulting field $E^{(+)}(\hat R,t)$ is
given by 
\be
E^{(+)}(\hat R,t)=E_a \sin (k_a \hat R) e^{-i (\omega_0 -\nu)t -i \phi_a} +
E_b \sin (k_b \hat R) e^{-i (\omega_0 +\nu)t -i \phi_b} + E_0 e^ {-i
  k_0 \hat R } e^{-i\omega _0 t-i\phi _0} \,\, .
\ee
The Lamb-Dicke parameter $\eta_j$ is defined by $k_j \hat R \equiv \eta_j (a
+a^\dagger)$, where $ \eta_j = \pi a_0 / \la_j$ with $a_0$ being the amplitude of the
ground state of the trap potential and $\la_j$ the optical
wavelength. Attention is confined here to the experimentally viable Lamb-Dicke
limit $\eta_j \ll 1$ and hence the Hamiltonian can be expanded to the first order
in $\eta_j$.
Let us define a rotating frame given by the unitary transformation
$\rho^\prime=U^\dagger \rho \, U$ where $U=\exp[-i(\nu a^\dagger 
a + \omega _0 \sigma _z /2)t]$ and $\rho$ is the density matrix of our system. 
By omitting the fast oscillating terms by assuming that $\nu$ is much larger
than any other parameters characterizing the system, the master equation for
$\rho^\prime$ is given by 
\be
{ d \rho^\prime \over d t }= -i [ H^\prime,\rho^\prime ] \,\, ,
\ee
where
\be
H^\prime =\sigma_+ (g_a a
  + g_b a^\dagger) + { \Omega_0 \over 2} \sigma_+ + h.c. \,\, ,
\label{ham1}
\ee
where $a,a^\dagger$ are the annihilation and creation operators for the trap
motion, $\sigma_+ := |e 
\ran \lan g|$, $\sigma_- := |g \ran \lan e|$ and $\sigma_z := |e\ran \lan e| - 
|g \ran \lan g|$ are the Pauli matrices describing the two-level transition of
frequency $\omega_0$. For the above the following conditions were taken;
for $g_j=\eta_j 
\Omega_j e^{-i \phi_j} /2$ with $\Omega_j$ being the real Rabi frequency of the
$j$-th laser ($\phi_0=0$) it is assumed that $ \Omega_0 \ll \Omega _{1,2}$ so that
terms of order $\eta_0 \Omega_0$ can be neglected. 

In this Hamiltonian the excitation of the ion is coupled with two transitions
in the vibrational modes. The first, produced by laser $1$, is a lowering of the
vibrational number
from, for example $n$ to $n-1$, while the second, laser $2$, raises the vibrational
number from $n$ to $n+1$. Coherent application of these two processes establishes
Raman transitions between alternative vibrational states (i.e. between $|n-1 \ran$
and $|n+1\ran$ ), finally giving a nonlinear effect. It is interesting to note that 
performing these transitions without the intermediate atomic levels would
require quadratic terms in the creation and annihilation operators. 
Going back to a resonance frame by making the inverse transformation
$\tilde U= \exp[-i(\nu a^\dagger a +g \sigma _z /2)t]$ one obtains
\be
P\equiv S^\dagger (\epsilon) D^\dagger(\alpha) \rho D(\al) S(\epsilon) \,\, ,
\ee
satisfying $ d P / d t = -i [ H_{JC}, P]$ with
\be
H_{JC}= \nu a^\dagger a + {g \over 2} \sigma _z + g(\sigma _+ a \, e^{-i
  (g-\nu)t} +a^\dagger \sigma_- \, e^{i(g-\nu)t}) \,\, ,
\label{JC}
\ee
and $\epsilon =r_1 e^{i\theta_1}$, $r_1=-1/2\tanh^{-1}(|g_b|/|g_a|)$,
$\theta_1=\phi_a-\phi_b +2\nu t$, $\alpha=-\Omega_0/(2g)e^{i\nu t}$, $g=|g_a| \cosh
r_1 -|g_b| \sinh r_1$. 
For $g=\nu$, $H_{JC}$ is the Hamiltonian of the Jaynes-Cummings model
describing a two-level atom coupled to a harmonic oscillator. 

\section{Jaynes-Cummings Model and Degeneracy}

Holonomic quantum computation is based on cyclic adiabatic transformations which allow
exchange of populations between states with the same energy. Degeneracy of
energy thus has to be introduced in our system. The natural way to achieve
this with the JC model is to tune the parameter $g$ in such a way that two
eigenstates of the Hamiltonian become degenerate.
In the JC model derived in the foregoing the Hamiltonian
has the ground internal state $|g\ran$ with energy $\omega_1=-g /2$ and the
excited state $|e\ran$ with energy $\omega_2=g /2$.
The lowest energy eigenstate of $H_{JC}$ is $|g,0\ran =|g\ran
\otimes |0\ran$ given as the tensor product of the internal and the vibronic
ground states, and it has the energy $E_0=\hbar \omega_1$. The rest of the
eigenstates (dressed states) are
given, for the resonant case $\nu=\omega_2 -\omega_1=g$, by
\be
|n,\pm \ran= \sqrt{1 \over 2} \big( |g,n+1\ran \pm |e,n\ran\big) \,\, ,
\ee
with the eigenvalues $E_{n \pm} =\hbar(\omega _1 +\nu(n+1)  \pm \nu \sqrt{n+1})$,
respectively. The resonant condition $g=\nu$ also creates the desired
degenerate condition $E_{deg}=E_0=E_{0-}$ and hence a two-dimensional encoding
space is spanned by the orthogonal states
\be
\{|g,0\ran , |0,-\ran=\sqrt{1 \over 2}(
|g,1\ran - |e,0\ran) \} \equiv \{|i \ran \, / \,\, i=0,1\} \,\, ,
\label{qubit}
\ee
where $|0\ran$ and $|1\ran$ are the qubit states.
None of the rest of the dressed states can have eigenvalues the same as $E_{deg}$,
which makes the degeneracy strictly 
twofold. The energy of this subsystem is $E_{deg}=\hbar\omega_1$ and, in
contrast to the rest of $E_{n \pm}$, it does not 
depend on the trap frequency $\nu$. The immediately higher
energy state is the $|1-\ran$ state
with energy $E_{1-}=\hbar (\omega_1 +(2-\sqrt{2})\nu)$. The energy difference
between it and the degenerate states gives the energy scale with which adiabatic
changes have to be 
compared in order not to have mixing of our encoding states with the rest of the
dressed states. 

The condition $g=\nu$ for producing degeneracy has to be compared with the
derivation of the Hamiltonian (\ref{ham1}) where we omitted fast oscillating
terms with frequency $2\nu$ in relation to $g=\nu$. A more plausible 
condition can be obtained if the 
constructed degeneracy is between the states $|0,+\ran$ and $|1,-\ran$, where
the omitted terms oscillate with frequencies six times as large as the
terms depicted in Hamiltonian (\ref{ham1}). In this case simulations
showed that squeezing of the vibrational states can be faithfully
produced with maximum amplitude $r_1 \approx 2$. Even though the
results will be the same as those presented here, the mathematical
derivation is somewhat more complicated, and so attention 
will be restricted to the simple case of the degeneracy between
$|g,0\ran$ and $|0,-\ran$. 
In addition, having strong lasers may influence the internal
structure of the atoms due to the Autler-Townes splitting
\cite{Autler}. Here we consider the case of weak binding limit where
along the variation of the laser amplitudes the internal atomic levels
employed here remain unaffected.

In the previous section, it was seen that the states of the
Hamiltonian $H$ are the
squeezed and displaced states of $H_{JC}$. Hence, $H$ has a degenerate
subspace spanned by the rotated basis 
$\{|\alpha,\epsilon ; i \ran \equiv D(\al) S(\epsilon) | i \ran \,/ \,\,
i=0,1\}$. This allows the degenerate states to be displaced and squeezed at
will by varying the phases and 
amplitudes of the electric fields of the lasers. Note that the dressing of the
internal states with the vibrational modes provides us with full
controllability of the qubit encoded in the two-dimensional degenerate subspace
simply by manipulating the vibrational modes without employing transitions
between the internal states.

\section{One-qubit Holonomic Gates} \label{oneg}

At this point we have all the ingredients necessary for performing holonomic
quantum computation for one qubit. The mathematical steps now to be
performed are similar to those used to calculate the holonomies for
the optical setup \cite{hol3}. With $H_0$ being the JC Hamiltonian and ${\cal M}$
the parametric space spanned by the coordinates
$\{x,y,r_1,\theta_1 \}$, for $\al=x+iy$ and $\ep=r_1 e^{i\theta_1}$, we are able
to perform iso-spectral transformations belonging to the family ${\cal F}=\{H_\sigma={\cal
U}(\sigma)\,H_0\, {\cal U}^\dagger(\sigma)\,/\,\,\sigma\in{\cal M}\,;\,\,{\cal U}
(\sigma)=D(\al)S(\epsilon)\}$ of Hamiltonians unitarily 
equivalent to and therefore iso-spectral with $H_0$, where $H_0=H_{\sigma_0}$ for
some $\sigma_0 \in {\cal M}$. 
With the operator $\tilde A_\sigma ={\cal U}^\dagger {\p \over \p \sigma}
{\cal U}$ defined for ${\cal U}=D(\al)S(\epsilon)$, the holonomic connection $A_\sigma$
of our model has the matrix elements $ A_\sigma ^{ij}=\lan i | \tilde
A_\sigma |j \ran$. Their components are given explicitly in the following
formulas:
$
\tilde A_\al =a^\dagger \cosh 2 r_1 + a e^{-i \theta_1} \sinh 2 r_1 + {\bar \al
  \over 2}$, $\tilde A_{\bar \al} =- \tilde A^\dagger _ \al$,
$\tilde A_{r_1}= e^{i \theta_1} \left. a^\dagger \right.^2 -e^{-i\theta_1} a^2$,
$\tilde A_{\theta_1}= {i \over 4} (\cosh 4 r_1 -1) (2 a^\dagger a +1) + {i \over
  4} \sinh 4 r_1 (e^{i \theta _1 } \left. a^\dagger \right.^2 + e^{-i \theta _1 }
a^2)$, eventually yielding
\[ \begin{array}{cc}
A_x=A_\al+A_{\bar \al}=&  
 \left[  \begin{array}{ccc}     -iy & -{1 \over \sqrt{2}}(\cosh 2r_1 - e^{i\th_1} \sinh 2r_1) \\
                                {1 \over \sqrt{2}}(\cosh 2r_1-e^{-i\th_1} \sinh 2r_1)& -iy \\
\end{array} \right] \,\, ,
\end{array}\]

\[ \begin{array}{cc}
A_y=i(A_\al-A_{\bar \al})=&  
 \left[  \begin{array}{ccc}     ix & {i \over \sqrt{2}}(\cosh 2r_1 + e^{i\th_1} \sinh 2r_1) \\
                                {i \over \sqrt{2}}(\cosh 2r_1 + e^{-i\th_1} \sinh 2r_1) & ix \\
\end{array} \right] \,\, ,
\end{array}\]

\[ \begin{array}{ccc}
A_{r_1}=&  
 \left[  \begin{array}{ccc}    0 & 0 \\
                               0 & 0 \\
\end{array} \right] \,\,\,\,\,\, ,\,\,\,\,\,\,\,\,\,\,\,\,\,\,
A_{\th_1}=&  
 \left[  \begin{array}{ccc}    1 & 0 \\
                               0 & {3 \over 2} \\
\end{array} \right] {i \over 4} (\cosh 4 r_1 -1)  \,\, .
\end{array}\]

Let us briefly mention that these connection components, $A_\sigma$, give rise
to two particular field strength components, $F_{\sigma_1 \sigma_2}=\p_{\sigma_1} 
A_{\sigma_2}-\p_{\sigma_2}A_{\sigma_1}+[A_{\sigma_1},A_{\sigma_2}]$,
analytically given by
\be
F_{r_1 x}=-i\left[ \begin{array}{cc}    0 & -i \\
                                        i & 0 \\
\end{array} \right] \sqrt{2}e^{-2 r_1} \,\,\,\,\, ,\,\,\,\,\,\,\,\,\,\,\,\,\,\,\,\,\,\,\,
\text{for $\theta_1=0$} \,\, ,
\ee
and
\be
F_{r_1 y}=i \left[ \begin{array}{cc}    0 & 1 \\
                                        1 & 0 \\
\end{array} \right] \sqrt{2}e^{-2 r_1} \,\,\,\,\, ,\,\,\,\,\,\,\,\,\,\,\,\,\,\,\,\,\,\,\,
\text{for $\theta_1=\pi$} \,\, .
\ee
Holonomic gates can now be constructed by traversing closed paths in
${\cal M}$. In particular, we shall choose two convenient planes in the
four-dimensional control manifold and evaluate the holonomies generated by any loop lying
on those planes. Explicitly, the loop $C_I \in \left. (x,r_1)\right._{\th_1=0}$ gives
$\Gamma_A(C_I)=\exp -i\hat \sigma_1 \Sigma_I$,
with $\Sigma_I:=\int_{D(C_I)} \!dxdr_1 2 e^{-2r_1}$.
The loop $C_{II} \in \left. (y,r_1)\right._{\th_1=\pi}$ gives
$\Gamma_A(C_{II})=\exp i\hat \sigma_2 \Sigma_{II}$,
with area $\Sigma_{II}:=\int_{D(C_{II})} \! dydr_1 2 e^{-2r_1}$.
In the above, $D(C_\rho)$ with $\rho=I,II$ is the surface on the relevant 
sub-manifold $(\sigma_i,\sigma_j)$ of ${\cal M}$ whose boundary is the path
$C_\rho$. 
These two unitaries, $\Gamma_A(C_I)$ and $\Gamma_A(C_{II})$, are sufficient to
produce any one qubit gate.
The exponential damping feature of the field strength has been transported into the
parameter of the holonomic gates, $\Sigma_\rho$, making them resilient to control errors
for large values of the squeezing parameter, $r_1$. This point will be
elaborated in a following section.

\section{Two ions and two collective vibrational modes}

An $N$ qubit system can be realized with a chain of $N$ ions. The internal
degrees of freedom of each ion are represented by a two-level system, while
the external degrees of freedom are described by the collective modes of the
crystal. These are the modes which diagonalize the potential (assumed here to
be harmonic) and the Coulomb repulsion between the ions. For sufficiently cold
crystals the ions oscillate harmonically around their equilibrium
position. Each qubit is composed of a two-level system corresponding to the
internal two-level transition of the ion, together with a collective mode of
the motion. We produce a two-qubit gate by interactions between the collective
modes of each qubit. This interaction is realized with a two-mode squeezing
or displacing transformation. In the following, these transformations are
constructed for two ions. 

In the case of a chain of two ions the collective modes are the ``center of
mass'' mode, where the two ions oscillate in phase, and the ``stretching''
mode, where the two ions oscillate out of phase. Let us define qubit 1 as
the combination of the internal levels of ion 1 with the center of mass mode,
and qubit 2 as the two levels of ion 2 together with the stretching mode,
analogously defined as in eq. (\ref{qubit}). Let $R_i$ be the displacement of ion $i$
from its equilibrium position and let the oscillations of the ions described by
two harmonic oscillators of frequency $\nu_1$, $\nu_2=\sqrt{3} \nu_1$ for the
center of mass and the stretching modes, respectively. The quantized
oscillations are described by the operators $a_i$, $a^\dagger _i$, with
$i=1,2$, which are the annihilation and creation operators of the center of
mass and the stretching mode, respectively.

Let us consider the following Hamiltonian of two ions where
only the first is coupled with both collective vibronic modes via the
appropriately detuned lasers:
\be
H_1=\nu_1 a^\dagger_1 a_1 + \nu_2 a^\dagger_2 a_2 + {\omega^1_0 \over 2}
\sigma^1_z +{\omega^2_0 \over 2} \sigma^2_z - {d \over \hbar} [\sigma^1 _+
E^{(+)}(\hat R_1,t)+\sigma^1 _- E^{(-)}(\hat R_1,t) ] \,\, .
\ee
The laser field acts on the first ion only, but
couples it to both oscillating modes by the following configuration:
\be
E^{(+)}(\hat R_1,t)= E_a \sin (k_a \hat R_1) e^{-i (\omega_L -\nu_2 )t -i \phi_a}
+E_b \sin (k_b \hat R_1) e^{-i (\omega _L +\nu_1)t -i \phi_b} \,\, ,
\ee
with $k_j\hat R_1 = \eta _j^{(1)} (a_1 +a^\dagger _1)+ \eta^{(2)} _j (a_2
+a^\dagger _2)$, where index on the frequencies and on the creation
and annihilation operators distinguishes between the two collective modes and the
index on the Pauli matrices distinguishes between the two atoms. In the
Lamb-Dicke limit we are able to expand the trigonometric functions up to the first
order in both of the parameters $\eta^{(1)}$ and
$\eta^{(2)}$. By moving to the rotated frame defined by $\rho^\prime
={\cal U}^\dagger \rho \,\,{\cal U}$, where 
${\cal U}=\exp [-i( \nu_1 a^\dagger _1 a_1 +\nu_2 a^\dagger _2 a_2 +
{\omega^1_0 \over 2} \sigma^1_z +{\omega^2_0 \over 2} \sigma^2_z)t]$ we
neglect all fast oscillating terms. These are phase factors with rotating
frequencies $\nu_{1,2}$ or $\nu_1 \pm \nu_2$. The resulting
master equation is given by
\be
{d \rho'_1 \over dt }= -i [H'_1 ,\rho'_1] \,\, ,
\ee
with
\be
H'_1=g_a^{(2)} \sigma^1_+
a_2 + g_b^{(1)} \sigma^1_+ 
a^\dagger _1 +h.c. \,\, ,
\ee
where $g^{(i)}_j=\eta^{(i)}_j \Omega_j e^{-i \phi_j} /2$.
Acting on the second ion with the detuning of the lasers interchanged between
the two collective frequencies, we obtain the following Hamiltonian
\be
H'_2=g_a^{(1)} \sigma^2_+  a_1 + g_b^{(2)} \sigma^2_+
  a^\dagger _2 +h.c. \,\, .
\ee
If the two interactions described in $H_1$ and $H_2$ are turned on {\it
simultaneously} then back to a resonant frame a two-mode squeezed density matrix 
$P=M^\dagger (\zeta) \big(\rho_1 \otimes \rho_2 \big) M(\zeta)$ is produced
satisfying $d P /dt =-i [H_{Tot},P]$, where the JC Hamiltonian of the two atoms
is given by 
\be
H_{Tot} =\nu_1 a^\dagger_1 a_1
+ \nu_2 a^\dagger_2 a_2 + {g_{(1)} \over 2} \sigma^1_z +{g_{(2)} \over
  2} \sigma^2_z + \{ g_{(1)} \sigma_+^1 a_1 + g_{(2)} \sigma_+^2 a_2 +h.c. \}
\,\, ,
\ee
with $M(\zeta)=\exp (\zeta a^\dagger _1 a^\dagger_2-\bar \zeta a_1 a_2)$,
$\zeta $ being rotated in the interaction frame defined by ${\cal
U}$ and given by $\zeta =r_2 e^{i \theta_2}$, $r_2=-\tanh^{-1} (|g_b^{(1)}|
/ |g_a^{(2)}|)=-\tanh^{-1}(|g^{(2)}_b|/|g^{(1)}_a|)$,
$g_{(1)}=|g^{(1)}_a|\cosh r_2 - |g^{(2)}_b| \sinh r_2 $,
$g_{(2)}=|g^{(2)}_a|\cosh r_2 - |g^{(1)}_b| \sinh r_2$  and $\theta_2=-\phi_b
+(\nu_1 +\nu_2)t$, where we have taken $\phi_a=0$. In
the same way, by interchanging the actions of the two laser fields on
the two atoms it is possible to obtain two-mode squeezing and two-mode
displacing transformations, the latter being defined by the unitary operator
$N(\xi)=\exp ( \xi a^\dagger_1 a _2 -\bar \xi a_1 a^\dagger _2)$. 

\section{Two-qubit Holonomic gates}

If the ground states of the two atoms $|g\ran_1$ and $|g\ran_2$ are kept on
the same energy level $\omega^1_1=\omega^2_1$, then the two degenerate spaces
of the two JC models, created by the conditions $g_{(1)}=\nu_1$ and
$g_{(2)}=\nu_2$, are also degenerate between each other. Hence, any
adiabatic evolution will allow the mixing of these four degenerate states
while they are kept isolated from the rest state space. The operators $\tilde
A_{r_2}=M^\dagger(\zeta) { \p \over \p r_2} M(\zeta)$ and
$\tilde A_{r_3} =M^\dagger (\zeta) \Big(N^\dagger (\xi) {\p \over \p r_3}
N(\xi) \Big) M(\zeta)$ give rise to the connection components $A^{ij} _\sigma
=\lan i| \tilde A_\sigma |j \ran$ for
$|i\ran$ and $|j\ran$ belonging to the tensor product basis of the two qubits
$|1\ran=|00\ran$, $|2\ran=|01\ran$, $|3\ran=|10\ran$ and
$|4\ran=|11\ran$. Hence, we obtain
\bq
&&
\tilde A_{r_2} =e^{i \theta_2} a^\dagger _1 a^\dagger _2 - e^{-i \theta_2} a_1
a_2 \,\, ,
\no \no
&&
\tilde A_{r_3} =(e^{i \theta_3} a^\dagger _1 a_2 -e^{-i \theta_3} a_1
a^\dagger _2) (\cosh^2 r_2 + \sinh ^2 r_2) +
\no \no
&&
\,\,\,\,\,\,\,\,\,e^{i\theta_3} (e^{i\theta_2}\left. a^\dagger_1 \right.^2 +
e^{-i \theta_2}\left. a^\dagger_2 \right.^2 ) \sinh r_2 \cosh r_2 - 
e^{-i\theta_3} (e^{-i\theta_2} a_1^2 + e^{-i \theta_2} a_2^2 ) \sinh r_2 \cosh
r_2 \,\, ,
\eq

\[ \begin{array}{ccc}
A_{r_2}=&  
 \left[  \begin{array}{cccc}    0 & 0 & 0 & -e^{-i\th_2} \\
                                0 & 0 & 0 & 0 \\
                                0 & 0 & 0 & 0 \\
                                e^{i\th_2} & 0 & 0 & 0 \\    
\end{array} \right] \sqrt{1 \over 2} \,\,\,\,\,\, ,\,\,\,\,\,\,\,\,\,\,\,\,\,\,
A_{r_3}=&  
 \left[  \begin{array}{cccc}    0 & 0 & 0 & 0 \\
                                0 & 0 & -e^{-i\th_3} & 0 \\
                                0 & e^{i\th_3} & 0 & 0 \\
                                0 & 0 & 0 & 0 \\    
\end{array} \right] \sqrt{{1 \over 2}} (2 \cosh ^2 r_2 -1) \,\, .
\end{array}\]
The relevant field strength component is given by
\[\begin{array}{ccc} 
F_{r_2 r_3}=&  
 \left[  \begin{array}{cccc}    0 & 0 & 0 & 0 \\
                                0 & 0 & -e^{-i\th_3} & 0 \\
                                0 & e^{i\th_3} & 0 & 0 \\
                                0 & 0 & 0 & 0 \\    
\end{array} \right] \sqrt{ 2}  \sinh 2r_2 \,\, .
\end{array}\]
Two-qubit holonomic gates can be produced by performing the following cyclic
evolutions. 
By means of a loop $C_{III} \in \left. (r_2,r_3)\right._{\th_2=\th_3=0}$ we obtain
$
\Gamma_A(C_{III})=\exp -i\hat \sigma_2^{(12)} \Sigma_{III} \,\, ,
$
with $\Sigma_{III}:=\int_{D(C_{III})} \! dr_2dr_3 2 \sinh 2 r_2$.
For $C_{IV} \in \left. (r_2,r_3)\right._{\th_2=0, \th_3=3\pi/2}$
the following holonomy is generated:
$
\Gamma_A(C_{IV})=\exp -i\hat \sigma_1^{(12)} \Sigma_{IV} \,\, ,
$
with the area given by $\Sigma_{IV}:=\int_{D(C_{IV})} \! dr_2dr_3 2 \sinh 2 r_2$.
In the above we have used
\bq
\hat \sigma _2^{(12)}:=
\left[  \begin{array}{cccc}    0 & 0 & 0 & 0 \\
                                0 & 0 & -i & 0 \\
                                0 & i & 0 & 0 \\
                                0 & 0 & 0 & 0 \\
\end{array} \right]
\nonumber
%\eq
\,\,\,\,\,\,\,\,\,\,\,\,\text{and}\,\,\,\,\,\,\,\,\,\,\,\,
%\bq
\hat \sigma _1^{(12)}:=
\left[  \begin{array}{cccc}    0 & 0 & 0 & 0 \\
                                0 & 0 & 1 & 0 \\
                                0 & 1 & 0 & 0 \\
                                0 & 0 & 0 & 0 \\
\end{array} \right] \,\, .
\nonumber
\eq
As an example, $\Gamma_A(C_{IV})$ with 
$\Sigma_{IV}=\pi/4$ gives the following non-trivial two-qubit gate:
\bq
U= {\sqrt{1 \over 2}}
\left[  \begin{array}{cccc}     \sqrt{2} & 0 & 0 & 0 \\
                                0 & 1 & -i\,\, & 0 \\
                                0 & -i\,\, & 1 & 0 \\
                                0 & 0 & 0 & \sqrt{2} \\
\end{array} \right] \,\, ,
\nonumber
\eq
which together with the holonomies produced by loops $C_I$, $C_{II}$ 
compose a universal set of transformations \cite{Loss}.

\section{Measurement Procedure}

At the end of the holonomic procedures it should be possible to distinguish
between the states $|g,0\ran$ and $|0,-\ran$ by a measurement. For this
purpose, a $\pi /2$ pulse is applied to the ion:
\be
U(\phi)=\exp \big[-i {\pi \over 4} (|e \ran\lan g| a e^{-i \phi}+h.c. )\big]
\,\, .
\ee
This evolution operator will not affect the state $|g,0\ran$, but only $|0,-\ran$.
In the rotated frame we are working in, the phase $\phi$ is shifted to $\tilde
\phi =\phi - (\omega_0 -g)t$. Hence, the state $|0,-\ran$ in the laboratory
frame becomes $-e^{-i {\omega_0 -g \over 2 } t} |e,0\ran$ when the timing is such
that $\tilde \phi = -\pi /2$. Direct measurement of the atomic energy levels
distinguishes between the two logical states of the qubits.  

\section{Control Error Resilience}

\begin{center}
\begin{figure}[ht]
\centerline{
\put(0,150){$x$}
\put(135,25){$r_1$}
\put(125,120){$C$}
\put(65,10){$(a)$}
  \epsffile{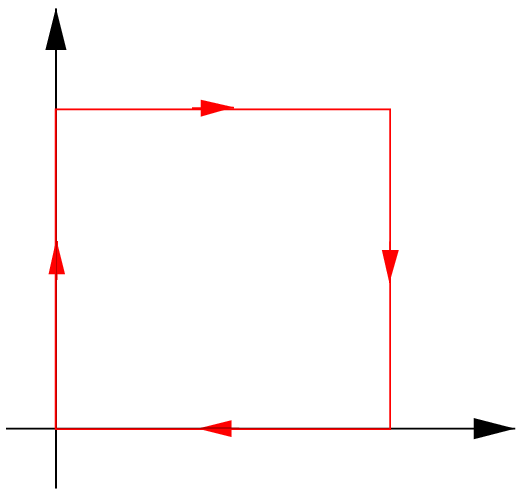}
\put(90,205){$x$}
\put(145,150){$y$}
\put(70,210){(1)}
\put(170,210){(2)}
\put(170,105){(3)}
\put(70,105){(4)}
\put(130,10){$(b)$}
\put(-10,90){{\boldmath $\Longleftrightarrow$}}
  \epsffile{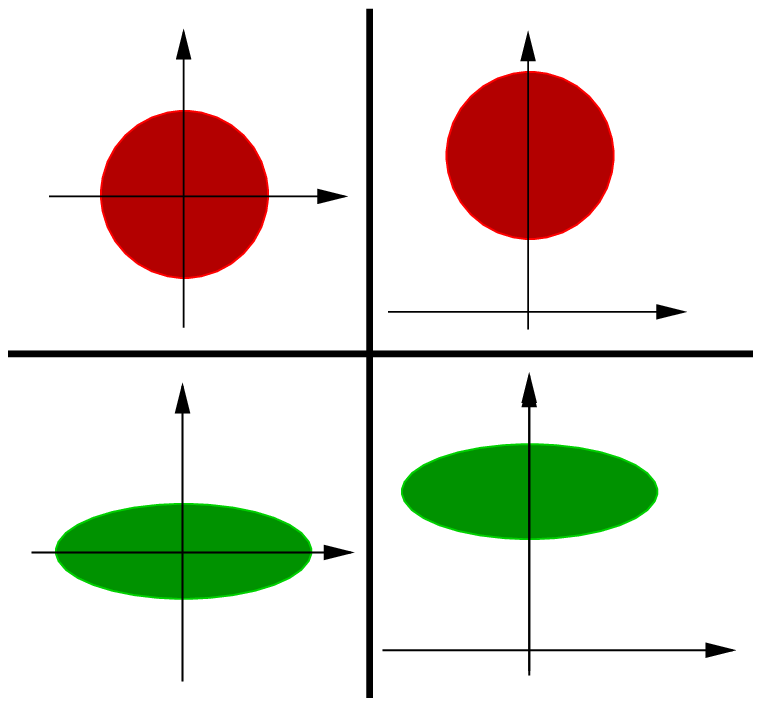}
}
  \caption[contour]{\label{loop1}
(a) The rectangular loop $C$ on the plane of the displacing
    parameter $x$ (or $y$) and the squeezing amplitude $r_1$. (b) The
    cyclic evolution in the phase space $(x,y)$: in (1) the $|0\ran$
    Fock state is depicted; then it is displaced along the $x$ axis in (2),
    squeezed (3), displaced back to the origin (4) and then, to complete
    the cyclic evolution, it is squeezed back to the original state $|0\ran$.
           }
\end{figure}
\end{center}

\begin{center}
\begin{figure}[ht]
\centerline{
\put(70,440){$\Delta \Sigma$}
\put(130,355){$\varepsilon_1$}
\put(250,350){$\varepsilon_2$}
  \epsffile{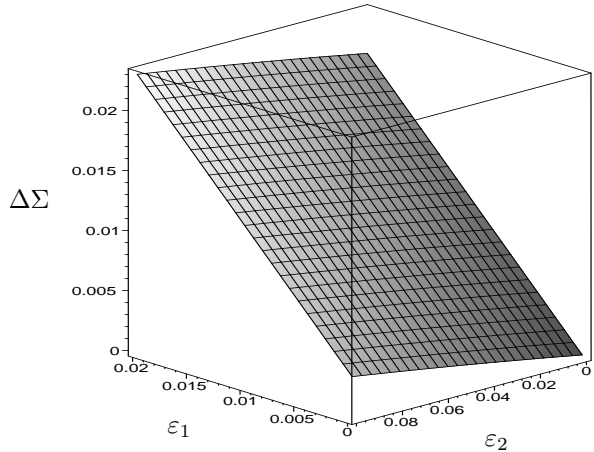}
}
\vspace{-12cm}
  \caption[contour]{\label{error1}
The error in the gate parameter, $\Delta \Sigma$, as a function of the
           displacing ($\varepsilon_1=\Delta x$) and squeezing
           ($\varepsilon_2=\Delta r_1$) 
           errors. The latter has a much smaller (about two
           orders of magnitude) influence on $\Delta \Sigma$. The dimensions
           of the rectangle are taken to be $x=1$ and $r_1=2$.
           }
\end{figure}
\end{center}

For the one-qubit gates presented in Section \ref{oneg} the
weight factor of the surface integrals $\Sigma_I$ and $\Sigma_{II}$ is $
e^{-2r_1}$. Let us introduce errors in the displacing 
and squeezing parameters. For the rectangular loop of Fig. \ref{loop1}(a) we may
take the borders, instead of being positioned at $(x,r_1)$, to be at
$(x+\varepsilon_1,r_1+\varepsilon_2)$ for small errors $\varepsilon_1$ and
$\varepsilon_2$.
Then the areas $\Sigma_I$ and $\Sigma_{II}$ are varied by 
\be
\Delta \Sigma= \varepsilon_1 \Big(1-e^{-2 (r_1+\varepsilon_2) }\Big) +x
e^{-2 r_1} \Big( 1- e^{-2 \varepsilon_2} \Big) \,\, ,
\ee
where its dependence with respect to both errors is given in
Fig. \ref{error1}. For values of $r_1 \approx 2$ the error $\varepsilon_2$ is
exponentially suppressed in relation to the contribution of
$\varepsilon_1$, which is linear. 

The resilience to the squeezing error is of the following nature. As has
been seen, the gate parameter can be viewed as the flux of the field strength
$F$ (which can be regarded as a virtual magnetic field). Its amplitude decreases for
large $r_1$. Hence, a loop which has one boundary at a large value of $r_1$
will include the same flux for small deformations of its boundary owing to the
exponential dumping of the field strength amplitude. This characteristic is
conceptually very interesting. A non-trivial bounded geometry,
i.e. one whose curvature decreases 
exponentially from a fixed point, looks like topology from far away. A simple
analogy can be drawn with the Aharonov Bohm effect. There, a magnetic field is
bounded in a solenoid and a charged particle travels around it. If the particle
performs a loop inside the solenoid, it acquires a phase given by the
enclosed flux of the magnetic field. This has a one-to-one correspondence with the
geometrical phases. But if the particle travels {\it far away} from the solenoid, it
acquires a fixed amount of phase, varying only by the integer number of
circulations the particle makes, which is a topological variable.

%\vspace{-1cm}

\section{Conclusions}

In this article, we have constructed an ionic setup where qubits are encoded in
the space of degenerate dressed states constructed from internal energy
levels of the atoms combined with their vibrational modes. The logical gates
are represented by the holonomies acting on the degenerate states. They are
produced by adiabatic
cyclic evolutions of the oscillating motion of the ions. These holonomies are
nonlinear functions of the parameters the experimenter is controlling.
This provides
a wide range of possibilities for constructing gates resilient to
control errors. In particular, we have seen one-qubit gates which afford 
exponential suppression of errors.

For two qubits the two-mode squeezings and displacings produce holonomies with
hyperbolic dependence on the control parameters. It is possible to use
two-mode control manipulations of higher order in the creation and annihilation
operators than the bilinear forms we have seen with
$M(\zeta)$ and $N(\xi)$. Even though their transformations have not
been mathematically studied sufficiently to evaluate their
holonomies, they appear promising for producing trigonometric \cite{Mista} or
even experimentally decreasing functions of the control parameters, while they
are experimentally easy to construct \cite{Steinbach}.

For quantum computation the threshold of error rate before which error correction
can be efficiently applied is of the order of $10^{-4}$. This very low limit
is more likely to be achieved by advantageous manipulation strategies rather
than by just trying by brute force to improve the controllability of our external
parameters. Holonomies provide the advantage that the parameters of our gates
are engineered functions of the experimental parameters. Furthermore, they are
resilient to statistical errors, thus providing overall an appealing framework for
quantum computation.

\vspace{1cm}
{\bf Acknowledgments} \hspace{0.1cm}
We would like to thank H. Walther and W. Lange
for inspiring conversations and for critical reading of the manuscript.

%%%%%%%%%%%%%%%%%%%%%%%%%%%%%%%%%%%%%%%%%%%%%%%%%%%%%%%%%%%%%%%%%%%%%%%%%%%

%%%%%%%%%%%%%%%%%%%%%%%%%%
%%%%%%%%%%%%%%%%%%%%%%%%%%%%%%%
%\end{multicols}
\end{document}